\documentclass[twocolumn,aps,prl,superscriptaddress,amssymb,showpacs]{revtex4}
\usepackage[dvips,final]{graphicx}
\begin{document}
\title{Quantum Andreev Oscillations
in normal-superconducting-normal nanostructures} 
\author{P. R\"odiger}\affiliation{Division of Superconductivity and Magnetism, Institut
f\"ur Experimentelle Physik II, Universit\"{a}t Leipzig,
Linn\'{e}stra{\ss}e 5, D-04103 Leipzig, Germany}
\author{P. Esquinazi}\email{esquin@physik.uni-leipzig.de}
\affiliation{Division of Superconductivity and Magnetism, Institut
f\"ur Experimentelle Physik II, Universit\"{a}t Leipzig,
Linn\'{e}stra{\ss}e 5, D-04103 Leipzig, Germany}
\affiliation{CMAM, Universidad Aut\'onoma de Madrid, Cantoblanco,
E-28049 Madrid, Spain} \affiliation{Laboratorio de F\'isica de
Sistemas Peque\~nos y Nanotecnolog\'ia,
 Consejo Superior de Investigaciones Cient\'ificas, E-28006 Madrid, Spain}
\author{N. Garc\'ia}\email{nicolas.garcia@fsp.csic.es}
\affiliation{Laboratorio de F\'isica de Sistemas Peque\~nos y
Nanotecnolog\'ia,
 Consejo Superior de Investigaciones Cient\'ificas, E-28006 Madrid, Spain}
\begin{abstract}
We show that the voltage drop of specially prepared
normal-superconducting-normal nanostructures show quantum Andreev
oscillations as a function of  magnetic field or input current.
These oscillations are due to the interference of the electron
wave function between the normal parts of the structure that act
as reflective interfaces, i.e. our devices behave as a Fabry-Perot
interferometer for conduction electrons. The observed oscillations
and field periods are well explained by theory.
\end{abstract}
 \maketitle

The possibilities of exploiting quantum mechanical effects -- with
all the interferences and other phenomena occurring in real
nano-devices -- find new expectations that may lead to the
fabrication of small devices with applications in new fields of
technology as ballistic electronics and spintronics and flux
devices combining normal and superconducting materials. Earlier
work  in semiconductors and STM experiments demonstrate the
existence of quantum oscillations \cite{bin85,gun66}. Recently,
spin-polarized resonant tunneling in magnetic tunneling junctions
showed large changes in the magnetoresistance due to the
interference of the carrier wave function \cite{MTJ}. In this work
we seek after quantum mechanical interference effects in the
magnetoresistance
 using
normal-superconducting structures without tunneling. Because one
of the characteristics needed is ballistic transport our
experiments have to be done at low temperatures and the systems
should be designed to have Fermi wavelength of the order or larger
than their relevant size.

 Assume a strip with a lateral structure M1/M2/M1,
where M1 and M2 are two different materials with different Fermi
energies $E_F$ and lengths $L_1$ and $L_2$ and where the
electrical current passes through them. Because of the different
$E_F$'s between M1 and M2, the one particle potential  can be
described by a barrier $U_2$ of length $L_2$ in M2 that acts
 as  a potential well where the wave function
behaves coherently having multireflections, i.e. a kind of
Fabry-Perot interferometer for electrons, a problem studied
recently for the case of fluctuations in the magnetoresistance of
graphene \cite{gar07}. Inset in Fig.~1(a) shows the
one-dimensional geometry of the system with the barrier $U_2(x)$
depending on the potential drop between the two ends of the
trilayer. The transmittivity along such structure is
\cite{messiah,gar07}
\begin{equation}
 T(U,\alpha)= \frac{4\Delta E(\alpha) E(\alpha)}{4E(\alpha)\Delta E(\alpha) +
 U_2^2\sin^2(2\pi L\sqrt{\Delta E(\alpha)\frac{2m}{\hbar^2}})}.
\label{T}
\end{equation}
The parameters in (\ref{T}) are: $\Delta E(\alpha) =
E(\alpha)-U_2;~E(\alpha)=(E_F+U)\cos^2(\alpha)$ is the energy of
the incoming particles forming an angle $\alpha$ to the interface,
$m$ the electronic mass and $\hbar = h/2\pi$.  Applying a
potential to the trilayer the potential in M2 is
$U_2(x)=U_0-(U/L_2)(x-L_1)$ ($L_1 \eqslantless x \eqslantless L_1+
L_2$).  The solution to this problem can be obtained as
combination of Airy functions \cite{bin85,gun66}. A good
approximation is to take the trapezoidal rule, i.e. $U_2(x)
\approx U_0 - 0.5U$ and calculate $T(U)$ following (\ref{T}) as
shown in Fig.~1(a). Interference effects require ballistic
transport in the potential region M2 with conveniently flat
interfaces to avoid multiple reflections. We will use M2 as a
superconductor (S) or as normal metal (N) and vice versa for M1,
of appropriate lengths.

\begin{figure}[]
\begin{center}
\includegraphics[width=88mm]{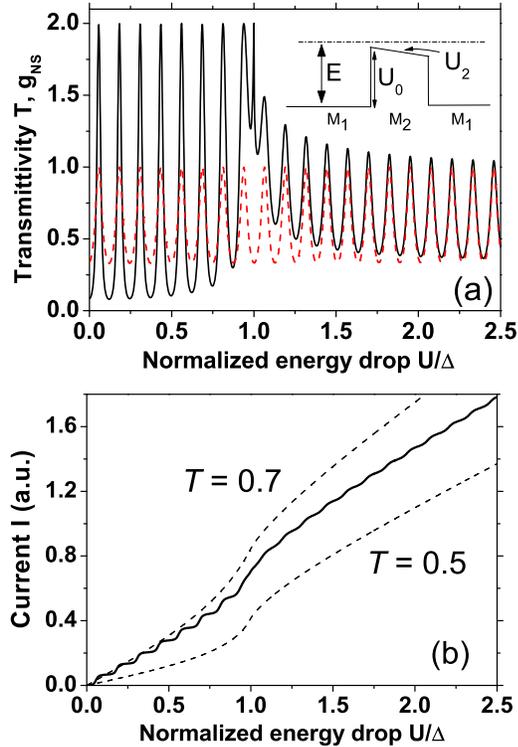}
\caption[]{(a) Inset shows the energy diagram for a M1/M2/M1
structure with a barrier $U_2(x)$ such that $U_2(L_1) = U_0$ (the
energy drop at M1 has been neglected for clarity).
 Main panel shows the transmittivity $T$ from (\protect\ref{T})
 for M2 in the normal state (dashed line)
 as well as the Andreev conductance
 $g_{\rm NS}$ (continuous line) vs. normalized energy drop $U/\Delta$ at M2.
 (b) Carrier current $I$ vs.
normalized energy drop  after (\protect\ref{I}). The two dotted
lines are calculated for constant transmittivity $T = 0.5, 0.7$
for a NS bilayer that shows no oscillations. The middle curve that
shows the oscillatory behavior is obtained with $T$ from
(\protect\ref{T}), see panel (a). Note that the strength of the
oscillations is larger for $U < \Delta$.} \label{theo}
\end{center}
\end{figure}

When M2 is a superconductor the current will be controlled by
Cooper pairs Andreev currents if $U<\Delta$ ($\Delta$ is the
superconducting gap) \cite{and64}. However at higher $U$ the
current is controlled by quasiparticles and for $U>3\Delta$ we
have practically conduction between normal materials
\cite{and64,blo82,gar88}. The solution for the current $I$ is in
this case:
\begin{eqnarray}
        I(U) &=&  A \int_0^U \left [ 1 - \frac{|(1-a^2)|^2 (1-
|T|)}{|1-a^2(1-|T|)|^2} \right. \nonumber \\
&&     \left.   + \frac{a^2|T|^2}{|1-a^2(1-|T|)|^2}  \right ] dU'
\,, \label{I}
\end{eqnarray}
where the term inside the integral is the Andreev conductance (in
units of quantum of conductance) $g_{\rm NS}(U,T)$
\cite{blo82,gar88}, which depends on $T$ in the {\it normal
state}; $a = (U/\Delta) - [(U/\Delta)^2 - 1]^{0.5}$; $A$ is a
constant that depends on the junction geometry and on the
integration average on $\alpha$. The solution for $I$ is depicted
in Fig.~1(b). Note that: (a) $g_{\rm NS}$ oscillates as a function
of the energy drop $U$ because of the resonances in $T(U)$ and it
tends to $T(U)$ when the product $U/\Delta \gg 1$, see Fig.~1(a).
(b) The measured voltage drop will oscillate as a function of $I$
and, at constant $I$, as a function of the magnetic field $B$
through  $U(B)/e \propto V(B) = IR(B)$, where $R(B)$ is the, in
general, non-oscillatory magnetoresistance of the sample.

The increment of the applied voltage drop $\Delta V$ needed to
obtain one oscillation is given by $\Delta V \sim \pi^2
(\hbar^2/m) (1/L_2^2)$. It can be seen that for $m$  of order of
the free mass the needed voltage change is $\Delta V \sim \mu$V to
mV for $L_2$ in the micro and nanometer range, respectively. For
example, if the superconducting gap is less that 1~meV with $L_2
\sim 10~$nm we need to have voltage drops of several millivolts,
i.e. the situation is practically a trilayer with normal
conducting materials. However, for $L_2 \sim 1~\mu$m, $\Delta V
\sim 1~\mu$V.

The NSN and SNS structures were produced combining electron beam
lithography with the deposition features of a FEI NanoLab XT 200
microscope. The tungsten precursor in the gas injection system
enables us to fabricate
  homogeneous, amorphous superconducting tungsten-carbide (WC)
  micro- and nano-structures by Ga$^+$-ion beam induced
deposition (IBID) with  critical temperature $T_c = 4 - 5~$K
\cite{spo07}. The deposition parameters were 30~kV accelerating
voltage, 10~pA Ga$^+$-ion current and 50\% overlap. All samples
were produced on $5 \times 5 \times 0.53$~mm$^3$ silicon
substrates with 150~nm insulating SiN layer. The lithographically
prepared multi-electrode structure was connected to a
multi-electrode chip and this was fixed in  a magneto-cryostat
system. The resistance measurements were done using four-wires
method with AC Linear Research LR700 bridge with a  multiplexer.
DC Current-voltage ($I-V$) characteristics as well as field $B$
dependent $V(B,I)$ curves were obtained using a Keithley 2182
nanovoltmeter with a 6221 current source.

\begin{figure}[]
\begin{center}
\includegraphics[width=88mm]{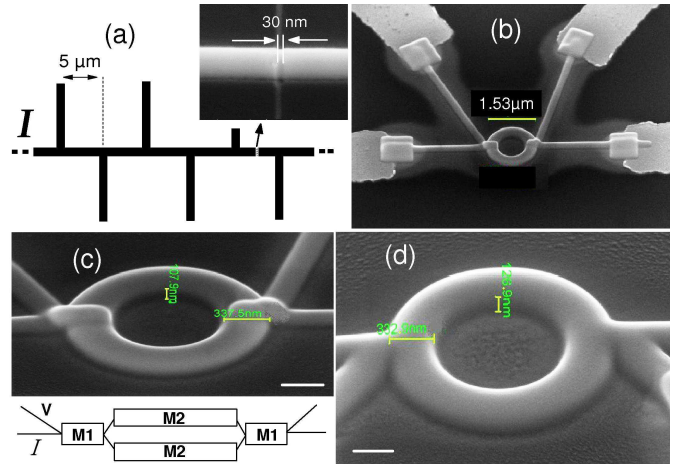}
\caption[]{(a) Sketch of the superconducting long strip with
voltages electrodes normal to it. The distance between them was
$5~\mu$m. The width and thickness were 200~nm and 80~nm. The SEM
picture at the upper right shows the region at and around one of
the slits. (b) SEM picture of the WC ring labeled WR$_2$ with the
contacts pads deposited after the ring. (c) A blow out of the same
ring as in (b). The cartoon below shows the equivalent circuit of
the ring with M1 and M2 the normal and superconducting paths. (d)
SEM picture of ring WR$_0$ produced in a single-step procedure.
The white bars in (c) and (d) indicate 300~nm distance.}
\label{photos}
\end{center}
\end{figure}

We will discuss first the results for SNS-type nanostructures. A
$\sim 60~\mu$m-long WC strip with eight voltage electrodes at a
distance of $5~\mu$m was fabricated in a single-step IBID
procedure enabling us to measure simultaneously different portions
of the strip (channels 1 to 8, see Fig.~\ref{photos}(a)). To
produce narrow, normal-conducting paths (M2) within the
superconducting strip with high-quality SN interfaces, minimizing
preparation time and costs, we irradiated it locally at the middle
of different voltage channels with the same Ga$^+$-ion beam
(30~kV, 1~pA, zero nominal thickness). In general, ion irradiation
strongly degrades the superconducting properties of a material,
see e.g. Ref.~\onlinecite{dyn}. The length of the slits was $L_2
\simeq 30~$nm, see Fig.~\ref{photos}(a). As reference some
channels were left without slits.  The sample W-SD-1 discussed
here showed a $T_c \simeq 4.3~$K defined at half-resistance at $B
= 0$~T. At input currents $I \lesssim 1~\mu$A and $B = 0$T the
slits short circuit the strip at $T < 4~$K due to the large
proximity effect as observed, for example, in Al thin films
\cite{kwo92}. At higher currents or fields the slits show normal
conducting behavior.

\begin{figure}[]
\begin{center}
\includegraphics[width=88mm]{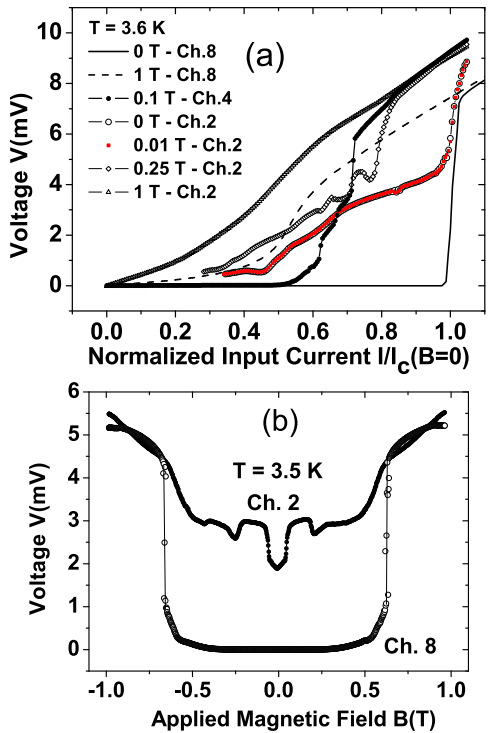}
\caption[]{(a) Voltage $(V)$ vs. input current ($I$) normalized by
the critical current $I_c(0) = 19.1~\mu$A at zero field at 3.6~K,
measured for two different channels, each one with one slit: Ch.~2
at 0~T$ \leqslant B \leqslant 1~$T and Ch.~4 at 0.1~T applied
field. Two curves were obtained for Ch.~8 (without irradiation,
$I_c(0) = 16~\mu$A) at 0~T and 8~T of the same strip sample. (b)
Voltage vs. applied field for Chs. 2 and 8 at 3.5~K and for an
input current of $12~\mu$A.} \label{ui}
\end{center}
\end{figure}

Figure~\ref{ui}(a) shows the voltage vs. current measured for
Chs.~2, 4 and 8 at 3.6~K at different magnetic fields applied
normal to the input current. At zero field and at $I < 16~\mu$A
the voltage drop in the homogeneous, unirradiated Ch.~8 is below
the 1~nV resolution. At the critical current $I_c \simeq 16~\mu$A
the voltage jumps to a value close to that expected for that
conducting path in the normal state, which has a resistance of
$\simeq 500~\Omega$. Within experimental resolution this curve
does remain without changes for an applied field of 0.01~T. At $B
> 0.01$~T the critical current decreases and flux-flow behavior is
observed. As example, Fig.~\ref{ui}(a) shows the data obtained at
$B = 1~$T. No hysteretic behavior was observed, which rules out
heating effects. Channels 2 and 4 show the voltage across two
different SNS structures, each with a single slit.   At $B = 0~T$
and 0.01~T the voltage drop below $I_c(0)$ is several orders of
magnitude larger than that obtained for Ch.~8 (as example we show
the signal of Ch.~2 in Fig.~\ref{ui}(a)). At fields below 1~T and
at $T = 3.6~$K Chs. 2 and 4 show clear oscillatory behavior
similar to that shown in Fig.~\ref{theo}(b), in contrast to that
of Ch.~8. At higher fields ($B \gtrsim 1~$T) the flux-flow
contribution prevails and the curves for all channels  resemble
qualitatively (Chs.~2 and 8 response is shown in
Fig.~\ref{ui}(a)). The quantum mechanical resonances can be also
observed in $V(B)$ at constant current, see Fig.~\ref{ui}(b),
because the magnetic field changes the voltage drop at the normal
conducting path due to its intrinsic magnetoresistance changing
therefore $T(U)$. The oscillation amplitude is of the order of mV,
i.e. larger than the superconducting gap $\Delta < 0.4~$meV, in
agreement with the model if $L_2 \simeq 30$~nm. Note that the free
electron gas model used here provides the general physics of the
phenomenon. More detailed analysis  of the band structure of M1
and M2 can explain the delicate structure of the experimental
curves.

\begin{figure}[]
\begin{center}
\includegraphics[width=88mm]{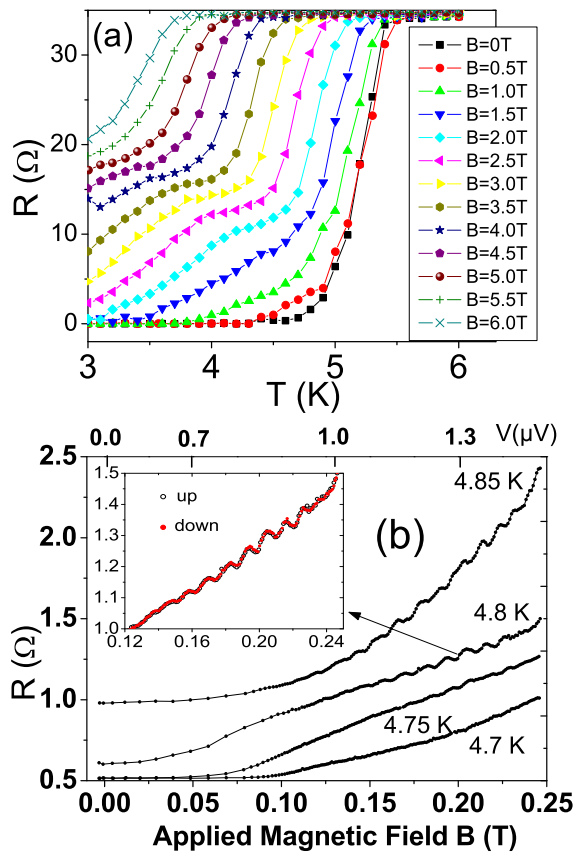}
\caption[]{(a) Resistance vs. temperature at different applied
fields of ring WR$_1$ measured with a current of $1~\mu$A, which
is 30 times smaller than $I_c$ at similar $B,T$. The critical
field of the superconducting transition of the ring itself (upper
transition) follows $B_{c2}(T) \simeq 8.1$[T]$ (1-(T/T_c)^2)$ with
$T_c \simeq 5.2~$K. (b) Resistance of the ring WR$_1$ vs. field at
different $T$. The inset blows out part of the 4.8~K curve
measured increasing and decreasing field. The dimensions of the
ring WR$_1$  were $(2.57, 1.6, 0.15)~\mu$m for the outer- and
inner-diameter and thickness, respectively.} \label{rtb}
\end{center}
\end{figure}

In what follows we discus NSN-type nanostructures with long
superconducting M2 paths. For reasons that will become clear below
and as an example of the deposition possibilities we prepared NSN
ring structures, see Fig.~\ref{photos}(b-d). To do the normal M1
paths we used a two-step IBID procedure, which consists in
re-deposit WC material on the top of the already deposited
superconducting ring path as the electrical contacts of the rings
WR$_1$ and WR$_2$. In this case the material M1, between the
superconducting electrodes and ring, shows a lower $T_c$ as the
resistance transition indicates, see Fig.~\ref{rtb}(a). The $T-$
and $B-$dependent measurements reveal that we have normal
(semi)conducting interfaces between the electrodes and the ring
structure at $T \gtrsim 4.7~$K at $B = 0~$T. To check that the
observed two-step transition as well as the oscillations described
below are not related to the ring structure itself, a one-piece
(e.g. one-step procedure) ring WR$_0$ was prepared in which the
contacts and ring are produced in a single-step IBID using
especial bitmap pattern generator features, see
Fig.~\ref{photos}(d).

Because in our NSN ring structures $L_2 \sim 3~\mu$m, we expect
oscillations in the submicrovolt range. In agreement with the
model these oscillations are nicely observed between $\sim 4.9~$K
and 4.7~K and around 0.2~T, see Fig.~\ref{rtb}(b). The field was
always applied normal to the ring area.
No oscillations were observed at lower temperatures, below the
critical line of the M1 material, above the critical line of the
ring as well as for the  homogeneous ring WR$_0$ in the whole
$T,B$-range. Note that the model \cite{gar07} applies only at
$T=0$K. However, the equivalent maximum energy of the oscillations
in the ring structure is $\sim 50~$neV much smaller than the
thermal energy of 0.4~meV. This indicates that the conducting path
M1 is a narrow band semiconductor because only in this case the
temperature excitation of electrons, intraband or to other
conduction band, is small.

We expect that the field period of the oscillations should be very
sensitive to the properties of the normal M1 part of the NSN
structure and therefore it should change from sample to sample.
This is indeed the case. The ring sample WR$_2$ shows a field
frequency of 0.1~T, about ten times larger than for sample WR$_1$,
implying that the M1 material should have a larger $E_F$ according
to the model.

The highly reversible behavior  of the oscillations at all
temperatures and fields (an example at 4.8~K is shown in the inset
of Fig.~\ref{rtb}(b)) does not support an interpretation of the
oscillations based on Josephson-like junction at the contact
positions. Magnetoresistance oscillations were observed
 in superconducting mesoscopic rings at very low
fields and explained by the oscillatory behavior of the critical
current density $I_c(B)$ at $T \sim T_c(B)$ \cite{mos93}.
Oscillations in the conductance were also observed in normal metal
rings with two tunnel junctions due to the magneto-electric
Aharonov-Bohm effect \cite{oud98}. In those works the field period
is related to the field needed to add one (or two) flux quantum in
the ring area, which for the geometry of our rings would give
$\sim 0.5$~mT - 2.5~mT,  far below the experimental observed value
of 12~mT - 100~mT for the rings WR$_1$ and WR$_2$, respectively.

In summary, in this work we showed the existence of quantum
Andreev oscillations in the magnetoresistance of NSN and SNS
nanostructures. These oscillations are observed also as a function
of magnetic field due to the intrinsic, non-oscillatory
magnetoresistance effect of the structure. The oscillations and
their field period are well explained by theory taking into
account ballistic transport and carrier wave function interference
within the structure. The effects here reported might be used to
study granular, superconducting materials, when neither Meissner
effect nor percolation can be observed. Similar oscillations as in
our NSN structures were recently observed in thin mesoscopic
structures of graphite \cite{esqui08}. This kind of quantum
mechanical resonances in the voltage should appear in every
nanostructure composed by different materials, if the electronic
conduction is ballistic and the roughness of the interfaces is
small enough.

We gratefully acknowledge the collaboration and support of K.
Schindler, J. Barzola-Quiquia, and M. Ziese. This work was done
with the support of a HBFG-grant no. 036-371, the Spanish CACyT
and Ministerio de Educaci\'on y Ciencia.


\begin{thebibliography}{14}
\expandafter\ifx\csname
natexlab\endcsname\relax\def\natexlab#1{#1}\fi
\expandafter\ifx\csname bibnamefont\endcsname\relax
  \def\bibnamefont#1{#1}\fi
\expandafter\ifx\csname bibfnamefont\endcsname\relax
  \def\bibfnamefont#1{#1}\fi
\expandafter\ifx\csname citenamefont\endcsname\relax
  \def\citenamefont#1{#1}\fi
\expandafter\ifx\csname url\endcsname\relax
  \def\url#1{\texttt{#1}}\fi
\expandafter\ifx\csname
urlprefix\endcsname\relax\def\urlprefix{URL }\fi
\providecommand{\bibinfo}[2]{#2}
\providecommand{\eprint}[2][]{\url{#2}}

\bibitem[{\citenamefont{Binnig et~al.}(1985)\citenamefont{Binnig, Frank, Fuchs,
  Garc\'ia, Reihl, Rohrer, Salvan, and Williams}}]{bin85}
\bibinfo{author}{\bibfnamefont{G.}~\bibnamefont{Binnig}},
  \bibinfo{author}{\bibfnamefont{K.~H.} \bibnamefont{Frank}},
  \bibinfo{author}{\bibfnamefont{H.}~\bibnamefont{Fuchs}},
  \bibinfo{author}{\bibfnamefont{N.}~\bibnamefont{Garc\'ia}},
  \bibinfo{author}{\bibfnamefont{B.}~\bibnamefont{Reihl}},
  \bibinfo{author}{\bibfnamefont{H.}~\bibnamefont{Rohrer}},
  \bibinfo{author}{\bibfnamefont{F.}~\bibnamefont{Salvan}}, \bibnamefont{and}
  \bibinfo{author}{\bibfnamefont{A.~R.} \bibnamefont{Williams}},
  \bibinfo{journal}{Phys. Rev. Lett.} \textbf{\bibinfo{volume}{55}},
  \bibinfo{pages}{991} (\bibinfo{year}{1985}).

\bibitem[{\citenamefont{Gundlach}(1966)}]{gun66}
\bibinfo{author}{\bibfnamefont{K.~H.} \bibnamefont{Gundlach}},
  \bibinfo{journal}{Solid State Electron.} \textbf{\bibinfo{volume}{9}},
  \bibinfo{pages}{949} (\bibinfo{year}{1966}).

\bibitem[{\citenamefont{Yuasa et~al.}(2002)\citenamefont{Yuasa, Nagahama, and
  Suzuki}}]{MTJ}
\bibinfo{author}{\bibfnamefont{S.}~\bibnamefont{Yuasa}},
  \bibinfo{author}{\bibfnamefont{T.}~\bibnamefont{Nagahama}}, \bibnamefont{and}
  \bibinfo{author}{\bibfnamefont{Y.}~\bibnamefont{Suzuki}},
  \bibinfo{journal}{Science} \textbf{\bibinfo{volume}{297}},
  \bibinfo{pages}{234} (\bibinfo{year}{2002}), \bibinfo{note}{{\rm C. W. Miller
  et al., Phys. Rev. Lett. {\bf 99}, 047206 (2007), T. Niizeki, N. Tezuka and
  K. Inomata, Phys. Rev. Lett. {\bf 100}, 047207 (2008)}}.

\bibitem[{\citenamefont{Garc\'ia}()}]{gar07}
\bibinfo{author}{\bibfnamefont{N.}~\bibnamefont{Garc\'ia}}, \bibinfo{note}{\rm
  arXiv:0706.0135}.

\bibitem[{\citenamefont{Messiah}(1999)}]{messiah}
\bibinfo{author}{\bibfnamefont{A.}~\bibnamefont{Messiah}},
  \emph{\bibinfo{title}{Quantum Mechanics}} (\bibinfo{publisher}{Dover Books on
  Physics}, \bibinfo{year}{1999}), ISBN \bibinfo{isbn}{0-486-40924-4}.

\bibitem[{\citenamefont{Andreev}(1964)}]{and64}
\bibinfo{author}{\bibfnamefont{A.~F.} \bibnamefont{Andreev}},
  \bibinfo{journal}{Sov. Phys. JETP} \textbf{\bibinfo{volume}{19}},
  \bibinfo{pages}{1228} (\bibinfo{year}{1964}).

\bibitem[{\citenamefont{Blonder et~al.}(1982)\citenamefont{Blonder, Tinkham,
  and Klapwijk}}]{blo82}
\bibinfo{author}{\bibfnamefont{G.~E.} \bibnamefont{Blonder}},
  \bibinfo{author}{\bibfnamefont{M.}~\bibnamefont{Tinkham}}, \bibnamefont{and}
  \bibinfo{author}{\bibfnamefont{T.~M.} \bibnamefont{Klapwijk}},
  \bibinfo{journal}{Phys. Rev. B} \textbf{\bibinfo{volume}{25}},
  \bibinfo{pages}{451} (\bibinfo{year}{1982}).

\bibitem[{\citenamefont{Garc\'ia et~al.}(1988)\citenamefont{Garc\'ia, Flores,
  and Guinea}}]{gar88}
\bibinfo{author}{\bibfnamefont{N.}~\bibnamefont{Garc\'ia}},
  \bibinfo{author}{\bibfnamefont{F.}~\bibnamefont{Flores}}, \bibnamefont{and}
  \bibinfo{author}{\bibfnamefont{F.}~\bibnamefont{Guinea}},
  \bibinfo{journal}{J. Vac. Sci. Technol. A} \textbf{\bibinfo{volume}{6}},
  \bibinfo{pages}{323} (\bibinfo{year}{1988}).

\bibitem[{\citenamefont{Spoddig et~al.}(2007)\citenamefont{Spoddig, Schindler,
  R{\"o}diger, Barzola-Quiquia, Fritsch, Mulders, and Esquinazi}}]{spo07}
\bibinfo{author}{\bibfnamefont{D.}~\bibnamefont{Spoddig}},
  \bibinfo{author}{\bibfnamefont{K.}~\bibnamefont{Schindler}},
  \bibinfo{author}{\bibfnamefont{P.}~\bibnamefont{R{\"o}diger}},
  \bibinfo{author}{\bibfnamefont{J.}~\bibnamefont{Barzola-Quiquia}},
  \bibinfo{author}{\bibfnamefont{K.}~\bibnamefont{Fritsch}},
  \bibinfo{author}{\bibfnamefont{H.}~\bibnamefont{Mulders}}, \bibnamefont{and}
  \bibinfo{author}{\bibfnamefont{P.}~\bibnamefont{Esquinazi}},
  \bibinfo{journal}{Nanotechnology} \textbf{\bibinfo{volume}{18}},
  \bibinfo{pages}{495202} (\bibinfo{year}{2007}).

\bibitem[{\citenamefont{Woods et~al.}(1988)\citenamefont{Woods, Katz,
  de~Andrade, Herrmann, Maple, and Dynes}}]{dyn}
\bibinfo{author}{\bibfnamefont{S.~I.} \bibnamefont{Woods}},
  \bibinfo{author}{\bibfnamefont{A.~S.} \bibnamefont{Katz}},
  \bibinfo{author}{\bibfnamefont{M.~C.} \bibnamefont{de~Andrade}},
  \bibinfo{author}{\bibfnamefont{J.}~\bibnamefont{Herrmann}},
  \bibinfo{author}{\bibfnamefont{M.~B.} \bibnamefont{Maple}}, \bibnamefont{and}
  \bibinfo{author}{\bibfnamefont{R.~C.} \bibnamefont{Dynes}},
  \bibinfo{journal}{Phys. Rev. B} \textbf{\bibinfo{volume}{58}},
  \bibinfo{pages}{8800} (\bibinfo{year}{1988}).

\bibitem[{\citenamefont{Kwong et~al.}(1992)\citenamefont{Kwong, Lin, Park,
  Isaacson, and Parpia}}]{kwo92}
\bibinfo{author}{\bibfnamefont{Y.}~\bibnamefont{Kwong}},
  \bibinfo{author}{\bibfnamefont{K.}~\bibnamefont{Lin}},
  \bibinfo{author}{\bibfnamefont{M.}~\bibnamefont{Park}},
  \bibinfo{author}{\bibfnamefont{M.}~\bibnamefont{Isaacson}}, \bibnamefont{and}
  \bibinfo{author}{\bibfnamefont{J.}~\bibnamefont{Parpia}},
  \bibinfo{journal}{Phys. Rev. B} \textbf{\bibinfo{volume}{45}},
  \bibinfo{pages}{9850} (\bibinfo{year}{1992}).

\bibitem[{\citenamefont{Moshchalkov et~al.}(1993)\citenamefont{Moshchalkov,
  Gielen, Dhall\'e, Haesendonck, and Bruynseraede}}]{mos93}
\bibinfo{author}{\bibfnamefont{V.~V.} \bibnamefont{Moshchalkov}},
  \bibinfo{author}{\bibfnamefont{L.}~\bibnamefont{Gielen}},
  \bibinfo{author}{\bibfnamefont{M.}~\bibnamefont{Dhall\'e}},
  \bibinfo{author}{\bibfnamefont{C.~V.} \bibnamefont{Haesendonck}},
  \bibnamefont{and}
  \bibinfo{author}{\bibfnamefont{Y.}~\bibnamefont{Bruynseraede}},
  \bibinfo{journal}{Nature} \textbf{\bibinfo{volume}{361}},
  \bibinfo{pages}{617} (\bibinfo{year}{1993}).

\bibitem[{\citenamefont{van Oudenaarden et~al.}(1998)\citenamefont{van
  Oudenaarden, Devoret, Nazarov, and Mooij}}]{oud98}
\bibinfo{author}{\bibfnamefont{A.}~\bibnamefont{van Oudenaarden}},
  \bibinfo{author}{\bibfnamefont{M.~H.} \bibnamefont{Devoret}},
  \bibinfo{author}{\bibfnamefont{Y.~V.} \bibnamefont{Nazarov}},
  \bibnamefont{and} \bibinfo{author}{\bibfnamefont{J.~E.} \bibnamefont{Mooij}},
  \bibinfo{journal}{Nature} \textbf{\bibinfo{volume}{391}},
  \bibinfo{pages}{768} (\bibinfo{year}{1998}).

\bibitem[{\citenamefont{Esquinazi et~al.}()\citenamefont{Esquinazi, Garc\'ia,
  Barzola-Quiquia, Mu{\~n}oz, R\"odiger, Schindler, Yao, and Ziese}}]{esqui08}
\bibinfo{author}{\bibfnamefont{P.}~\bibnamefont{Esquinazi}},
  \bibinfo{author}{\bibfnamefont{N.}~\bibnamefont{Garc\'ia}},
  \bibinfo{author}{\bibfnamefont{J.}~\bibnamefont{Barzola-Quiquia}},
  \bibinfo{author}{\bibfnamefont{M.}~\bibnamefont{Mu{\~n}oz}},
  \bibinfo{author}{\bibfnamefont{P.}~\bibnamefont{R\"odiger}},
  \bibinfo{author}{\bibfnamefont{K.}~\bibnamefont{Schindler}},
  \bibinfo{author}{\bibfnamefont{J.-L.} \bibnamefont{Yao}}, \bibnamefont{and}
  \bibinfo{author}{\bibfnamefont{M.}~\bibnamefont{Ziese}}, \bibinfo{note}{\rm
  arXiv:0711.3542}.

\end{thebibliography}

\end{document}